\documentclass[twocolumn,showpacs,aps,prb,superscriptaddress,floatfix]{revtex4}
\usepackage{graphicx}
\usepackage{dcolumn}
\usepackage{bm}

\begin{document}

\preprint{}

\newcommand{\tto}{Tb$_{2}$Ti$_{2}$O$_{7}$}
\author{K.C. Rule}
\affiliation{Department of Physics and Astronomy, McMaster University,
Hamilton, Ontario, L8S 4M1, Canada}
\author{J.P.C. Ruff}
\affiliation{Department of Physics and Astronomy, McMaster University,
Hamilton, Ontario, L8S 4M1, Canada}
\author{B.D. Gaulin}
\affiliation{Department of Physics and Astronomy, McMaster University,
Hamilton, Ontario, L8S 4M1, Canada} 
\affiliation{Canadian Institute for Advanced Research, 180 Dundas St. W.,
Toronto, Ontario, M5G 1Z8, Canada}
\author{S.R. Dunsiger}
\affiliation{Department of Physics and Astronomy, McMaster University,
Hamilton, Ontario, L8S 4M1, Canada}
\author{J.S. Gardner}
\affiliation {Brookhaven National Laboratory, Upton, New York 11973-5000, 
USA}
\affiliation{National Institute of Standards and Technology, 100 Bureau
Dr., Gaithersburg, MD, 20899-8562, U.S.A.}
\author{J.P. Clancy} 
\affiliation{Department of Physics and Astronomy, McMaster University, 
Hamilton, Ontario, L8S 4M1, Canada}
\author{M.J. Lewis}
\affiliation{Department of Physics and Astronomy, McMaster University,
Hamilton, Ontario, L8S 4M1, Canada}
\author{H.A. Dabkowska} 
\affiliation{Department of Physics and Astronomy, McMaster University, 
Hamilton, Ontario, L8S 4M1, Canada}
\author{I Mirebeau}
\affiliation{Laboratoire Leon Brillouin, CEA-CNRS, CE-Saclay, 91191
Gif-sur-Yvette, France}
\author{P. Manuel}
\affiliation{ISIS Facility, Rutherford Appleton
Laboratory, Didcot, Oxon, OX11 0QX, United Kingdom.}
\author{Y. Qiu}
\affiliation{National Institute of Standards and Technology, 100 Bureau 
Dr., Gaithersburg, MD, 20899-8562, U.S.A.} 
\affiliation{Department of Materials Science and Engineering, University 
of Maryland, College Park, MD, 20742, USA}
\author{J.R.D. Copley}
\affiliation{National Institute of Standards and Technology, 100 Bureau
Dr., Gaithersburg, MD, 20899-8562, U.S.A.}

\title{Field Induced Order and Spin Waves in the Pyrochlore 
Antiferromagnet \tto} 

\begin{abstract} 
High resolution time-of-flight
neutron scattering measurements on {\tto} reveal a rich low temperature
phase diagram in the presence of a magnetic field applied along [110].  
In zero field at T=0.4 K, {\tto} is a highly correlated cooperative
paramagnet with disordered spins residing on a pyrochlore lattice of 
corner-sharing tetrahedra.  Application of a small field condenses much of the 
magnetic diffuse scattering, characteristic of the disordered spins, 
into a new Bragg peak characteristic of a polarized paramagnet. At higher 
fields, a magnetically ordered phase is induced, which supports spin 
wave excitations indicative of continuous, rather than Ising-like spin degrees of freedom.

\end{abstract} 

\maketitle 
Geometrical frustration is a central theme in contemporary condensed 
matter physics, as it allows the possibility of exotic ground states 
\cite{Ramirez94}. Geometries which support magnetic frustration typically involve edge and 
corner shared triangles and tetrahedra, and these are common in nature.  
Pyrochlore magnets, with magnetic moments 
localized at the vertices of a cubic network of corner-sharing tetrahedra, 
have played a prominent role in this field.  Many such systems exist
with varied magnetic interactions and anisotropies expressed by 
real materials.

Idealized systems comprised of classical spins which interact on the
pyrochlore lattice are reasonably well understood.  
Antiferromagnetically-coupled Heisenberg spins are known to have
a disordered ground state on the pyrochlore lattice 
\cite{Reimers92, Moessner98, Harris94}.  Spins with 
local 
[111] Ising anisotropy, such that moments are constrained to point into or 
out of the tetrahedra, have been extensively studied for both 
antiferromagnetic
and ferromagnetic coupling \cite{denHertog00, Harris97}.  Interestingly, 
it is known that if the local anisotropy is
sufficiently strong, ferromagnetic exchange leads to a disordered ``spin 
ice"
ground state, while antiferromagnetic exchange gives rise to a {\bf Q}=0,
non-collinear N\'{e}el ordered state.

The rare-earth titanate pyrochlores, with the chemical composition 
A$_2$Ti$_2$O$_7$, where the A site is a trivalent rare-earth ion 
surrounded by a (distorted) cube of eight O$^{2-}$ ions and Ti is in its 
non-magnetic Ti$^{4+}$ state, have been of 
particular interest with regard to these latter calculations.  The crystal 
field states of rare-earth ions Ho$^{3+}$, Dy$^{3+}$, 
and Tb$^{3+}$ within A$_2$Ti$_2$O$_7$ are such that local [111] Ising anisotropy of the moments is 
expected \cite{Gingras00,Rosenkranz}.    As the exchange interactions 
are relatively weak, and the moment sizes are relatively large, the 
appropriate starting point Hamiltonian for a 
discussion of these materials is \cite{denHertog00}:

\begin{eqnarray}
{\cal H}&=&
\;\; -\sum_{\langle i,j\rangle}J_{ij}{\bf S}_{i}^{z_{i}}\cdot{\bf S}_{j}^{z_{j}}  \\
&+&  Dr_{{\rm nn}}^{3}\sum_{ \begin{array}{c} i>j \end{array}    }
\frac{{\bf S}_{i}^{z_{i}}\cdot{\bf
S}_{j}^{z_{j}}}{|{\bf R}_{ij}|^{3}} - \frac{3({\bf S}_{i}^{z_{i}}\cdot{\bf
R}_{ij}) ({\bf S}_{j}^{z_{j}}\cdot{\bf R}_{ij})}{|{\bf R}_{ij}|^{5}} \nonumber
\end{eqnarray} 
 
where \textit{z$_{i}$} denotes the local Ising axis.
Tb$^{3+}$ in {\tto} displays a ground state and 1st excited state doublet, 
separated by only about $\hbar\omega\sim$ 1.5 meV, with higher energy singlet states 
observed 
by inelastic neutron scattering at $\hbar\omega\sim$ 10.5 and 14.5 meV 
\cite{Gardner01}.  The moment 
associated with the ground state doublet is roughly 5 $\mu_B$ 
\cite{Gingras00}, but it 
is known to be sensitive to the precise oxygen coordination.  A 
perfect cube of O$^{2-}$ around the Tb$^{3+}$ gives rise, for example, 
to a non-magnetic doublet.

The magnetic 
ground state of {\tto}, at ambient pressure and in zero applied magnetic 
field, is known to remain a cooperative paramagnet to 
temperatures as low as 20 mK\cite{Gardner99, Gardner01, Gardner03}.  In some studies, 
indications of spin freezing, or ordering of an undetermined nature, have 
been observed at temperatures 
as high as 1 to 2 K \cite{Yasui02, Dunsiger03, Cv04}; 
however most of the magnetic spectral weight remains dynamic in 
frequency and diffuse in {\bf Q}-space to the lowest temperatures 
measured \cite{Gardner99, Gardner01, Gardner03}. 
The absence of order is quite enigmatic, as best estimates appropriate to {\tto} 
and arising from Eq. 1 suggest a Q=0 non-collinear N\'{e}el state 
below $\sim$ 1 K \cite{denHertog00, Gingras00}.

Earlier studies have shown that external perturbation by pressure and 
magnetic field may induce order in {\tto}.  Neutron studies on {\tto} 
in the presence of [111] magnetic fields show an increase to the 
scattering at nuclear allowed Bragg positions \cite{Yasui01}.  More 
recently, the application of hydrostatic pressure on powder samples, as 
well as combinations of hydrostatic and uniaxial pressure, in 
concert with applied magnetic fields on single crystals show the 
existence of a magnetically-ordered state at low temperatures and field\cite{Mirebeau05, MirebeauPF}.  
In addition, recent measurements on polycrystalline {\tto} 
in a relatively strong magnetic field shows evidence for very slow spin 
relaxation at temperatures as high as 10 - 20K \cite{Schiffer}. In 
this letter, we report new high resolution 
time-of-flight neutron measurements on a single crystal {\tto} in a 
magnetic field applied along a [110] direction.  These measurements reveal a  
complex magnetic field - temperature phase diagram, characterized by the well known 
cooperative paramagnet, a polarized paramagnet, and a high field long range ordered magnetic phase 
with accompanying spin wave excitations.

\begin{figure}
\centering
\includegraphics[width=8.5cm]{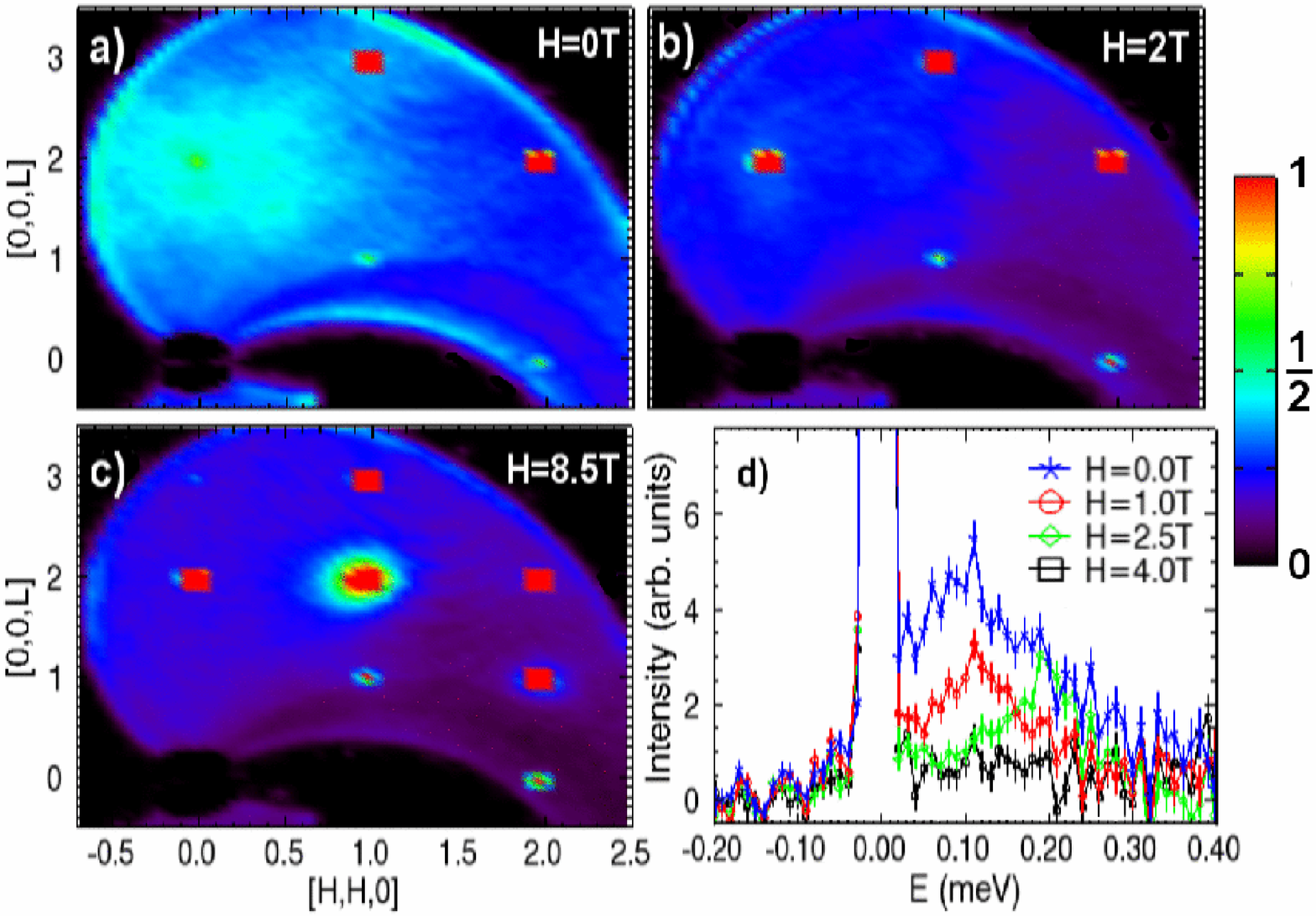}
\caption{Neutron Scattering for -0.5 meV $<$$\hbar\omega$$<$ 0.5 meV, and 
within 
the (H,H,L) plane of {\tto} at T=1 K is shown for (a) H=0 T, (b) H=2 T and 
(c) H=8.5 T.  Panel d) shows high energy resolution scattering at T=0.1 K, integrated 
in {\bf Q} and including the (0,0,2) position.  The 
quasi-elastic scattering extends to $\sim$ 0.25 meV and it is  
dramatically diminished in fields as low as 1 T.} 
\label{Figure 1} 
\end{figure}

\begin{figure}
\centering 
\includegraphics[width=8.0cm]{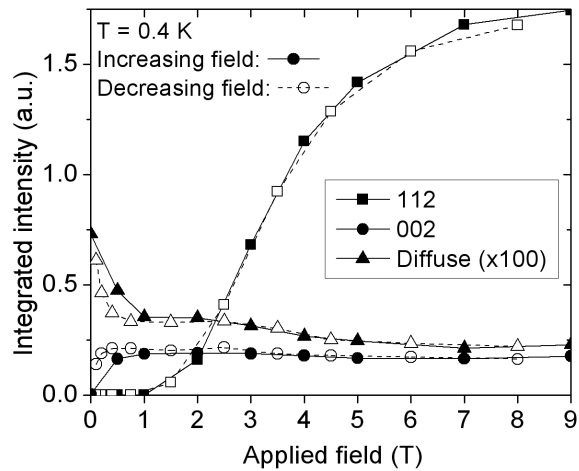}
\caption{The integrated intensity of the (1,1,2) and (0,0,2) Bragg peaks 
as a function of H at T=0.4 K are shown, along with the integrated 
quasi-elastic scattering making up the diffuse scattering.  Note the 
complementarity of the (0,0,2) Bragg scattering and the diffuse scattering 
and that both exhibit hysteresis with field.} 
\label{Figure 2.} 
\end{figure}

The single crystal sample was grown by floating zone image furnace
techniques and was cylindrical in shape with approximate dimensions of 2 
cm
long by 1 cm in diameter.  It is the same high quality single crystal
studied previously \cite{Gardner98}. Time-of-flight neutron scattering measurements 
were
performed using the Disk Chopper Spectrometer (DCS) at the
NIST Center for Neutron Research \cite{DCS}.  The DCS uses choppers to create 
pulses
of monochromatic neutrons whose energy transfers on scattering are
determined from their arrival times in the instrument's 913 detectors
located at scattering angles from -30 to 140 degrees. Measurements were
performed with 5 and 9 {\AA} incident neutrons.

Figure 1a, b, and c show reciprocal space maps in the (H,H,L) plane, 
integrating over -0.5$<$$\hbar\omega$$<$0.5 meV and taken at T= 1 K.  
These data employed 5 {\AA} incident neutrons.  Figure 1a shows data 
in zero applied field, while Figs 1b and c show data in H=2 T and 8.5 T 
respectively.  The data in Fig. 1a, in zero field, reproduce earlier 
neutron measurements \cite{Gardner01} of the ``checkerboard" diffuse scattering, 
characteristic of very short range spin correlations - over single 
tetrahedra, which is particularly pronounced around (0,0,2).  Nuclear 
Bragg peaks allowed by the $Fd\bar{3}m$ pyrochlore space group are easily observed 
at (3,1,1), (2,2,2), (1,1,1), and (2,2,0).  In addition a very weak Bragg 
peak is 
evident on close examination of the data at (0,0,2) which is {\it not} 
allowed by the $Fd\bar{3}m$ space group.  If nuclear in origin, its presence 
indicates {\tto} is not a perfect cubic pyrochlore.

Figures 1b and c show the appearance of Bragg peaks on application of a 
[110] magnetic field.  At 2 T, the map shows the diffuse 
scattering around (0,0,2) has largely disappeared, and a strong Bragg peak 
is 
evident at (0,0,2).  At higher fields there is clearly a transition to 
an ordered state, as the data at 8.5 T in Fig. 1c show the appearance 
of a set of new, intense Bragg peaks at most of the integer (H,H,L) 
positions in the field of view.

Measurements with 5 {\AA} incident neutrons and high resolution 
measurements using 9 {\AA} neutrons show the diffuse scattering around 
(0,0,2) to be quasi-elastic in nature with a characteristic extent in 
energy of $\sim$ 0.25 meV.  This is seen most clearly in Fig. 1d, which shows {\bf 
Q}-integrated scattering near (0,0,2) as a function of energy at 
T=0.1 K and several values of magnetic field applied along the [110] 
direction. 

Parametric studies of the Bragg features at (0,0,2) and (1,1,2), as well
as of the diffuse scattering near (0,0,2) as a function of field and
temperature were carried out to elucidate the new phase diagram for {\tto}
in a [110] magnetic field.  Figure 2 shows the integrated intensity of the
(0,0,2) and (1,1,2) Bragg peaks, and the diffuse scattering near (0,0,2)
at T=0.4 K. These data were acquired from a set of seven scans, comprising
a 3 degree sample rotation, and going through either the (0,0,2) or
(1,1,2) Bragg positions.  The (0,0,2)  integrated Bragg intensity rises
almost immediately from a very small value and saturates beyond $\sim$ 0.5
T.  The diffuse scattering shows the complementary behavior, with a rapid
fall off with increasing field, and then a leveling out at a small but
non-zero intensity.  At higher fields the diffuse scattering drops off
again, near 5 T, to background.  Measurements were made in both
increasing and decreasing magnetic field, and hysteresis is observed in
both the (0,0,2) Bragg and diffuse scattering intensity.

The (1,1,2) integrated Bragg intensity undergoes a sharp ``S-shaped" rise from zero 
starting near 2 T and saturates near 8 T.  No hysteresis is 
observed in its field dependence.  The diffuse scattering appears to 
decrease in intensity to zero as the (1,1,2) Bragg intensity goes 
through its inflection point near 5 T.   

Figure 3 shows the temperature dependence of both the (0,0,2) and (1,1,2)  
Bragg peaks in magnetic fields of 1 and 7 T.  As seen in Fig. 2, the
(1,1,2)  Bragg peak has zero intensity in a field of 1 T.  In an H=7 T
field its intensity falls off sharply with downwards curvature,
indicating a phase transition near T$_N$$\sim$ 3 K.  In striking contrast,
the (0,0,2) Bragg peak in an H=1 T field falls off with increasing
temperature, but remains non-zero to temperatures at least as high as 24 K.  The
upwards-curvature associated with the (0,0,2) T-dependence suggests no
phase transition occurs in this temperature range.  As can also be seen in Fig. 3, the
(0,0,2)  intensity in H=7 T undergoes an anomaly at T$_N$, but remains
large to at least 20 K.

The phase boundary between the high field ordered magnetic state and the
polarized paramagnetic state is shown in the inset of Fig. 3.  This line 
of phase transitions is identified from the time-of-flight neutron results 
of the form shown in Figs. 2 and 3, along with complementary triple axis 
neutron measurements and ac-susceptibility measurements\cite{Graeme}.

\begin{figure}
\centering
\includegraphics[width=8.5cm]{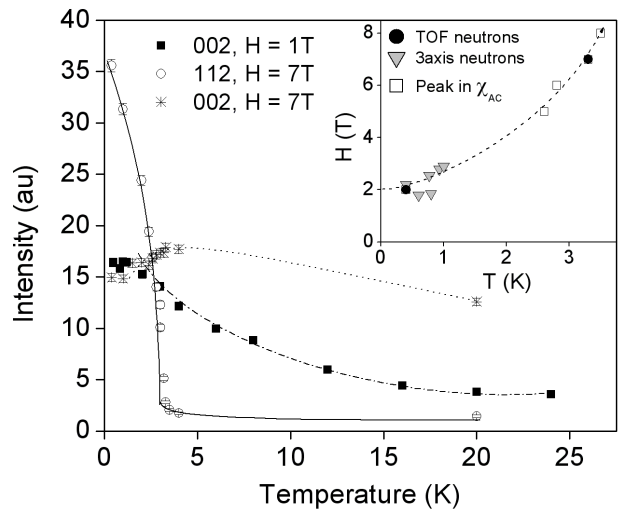}
\caption{The temperature dependence of the (1,1,2) Bragg peak in H=7 T, as 
well as the (0,0,2) Bragg peak in both H=1 T and 7 T is shown. The inset 
shows the phase diagram for the magnetically long range ordered 
state in {\tto}, as determined by time-of-flight and triple axis neutron 
scattering and by ac-susceptibility measurements. All lines shown are 
guides-to-the-eye.} \label{Figure 3} 
\end{figure}

Time-of-flight measurements taken at a single sample orientation within 
the (H,H,L) plane and integrating along (H,H,0) are shown in Fig. 4.  
These measurements, taken at T=0.4 K, approximate the inelastic 
scattering spectrum within the (0,0,L), -$\hbar\omega$ plane around (0,0,2).  
The four panels show data in Fig. 4a) zero magnetic field, b) H=1 T, c) 
H=2 
T, and d) H=3 T.  The zero field results show a quasi-elastic 
spectrum at energies less than $\sim$ 0.3 meV responsible for the diffuse 
scattering around (0,0,2) and a gap in the 0.3 - 0.8 meV region.  At 
higher energies we observe the same inelastic modes seen previously\cite{Gardner01}, 
a relatively broad distribution of inelastic scattering from 0.8 meV to 
1.8 meV, with a minimum in the dispersion around (0,0,2).  On application 
of an H=1 T field, sufficient to generate the full Bragg intensity 
at (0,0,2) (see Fig. 2), the inelastic scattering is qualitatively similar to 
that in zero field, although the quasi-elastic diffuse scattering appears 
weaker and the inelastic bands of scattering between 0.8 meV and 1.8 meV 
are somewhat narrower.

At H=2 and 3 T, shown in Fig. 4c) and d), the spectrum is qualitatively 
different than that at the lower applied fields.  Most strikingly, we 
observe sharp, dispersive spin wave excitations, which appear to have 
minima in their dispersion near (0,0,1) and (0,0,3).  In addition the 
quasielastic scattering that was responsible for the ``checkerboard" of 
diffuse scattering in zero field, is now clearly resolved as a relatively 
dispersionless inelastic mode at $\hbar\omega \sim$ 0.3 meV.  The higher 
energy inelastic scattering is resolved into relatively narrow energy 
bands.  As seen in Fig. 2, this new magnetic inelastic spectrum occurs 
within the magnetically ordered phase, characterized by the strong (1,1,2) 
Bragg peak.  

We identify the (0,0,2) Bragg peak, which 
appears on application of a very small applied field as a signature of a polarized 
paramagnet.  We do so based on two observations: i) the absence 
of a clear signature of a phase transition associated with the 
(0,0,2) intensity in an H=1 T field below 24 K;  and ii) the 
absence of collective spin wave excitations at H=1 T, as are observed at 
higher fields. 

The Tb sites in the pyrochlore structure can be thought of as lying on two
sets of chains oriented along orthogonal [1,1,0] directions.  The
application of a magnetic field along one particular [1,1,0] is expected
to polarize half of the Tb sites, such that moments on chains parallel to
the field direction point align along the local [1,1,1] direction with a
component parallel to the field. Such a polarized paramagnet displays
magnetic Bragg peaks at the (0,0,2) positions as well as the nuclear Bragg
positions of the $Fd\bar{3}m$ space group:  (3,1,1), (2,2,2), (1,1,1), and
(2,2,0), as observed experimentally in Fig. 1c).  This state is neither
expected to display a phase transition, nor to support spin waves, as the
long range correlations responsible for the magnetic Bragg peaks are due
to a single-ion canting of a subset of the moments along the applied field
direction.

In contrast, the high field magnetically ordered state shows both a phase
transition near T$_N$$\sim$ 3 K, and well defined spin wave excitations.  
The dispersive spin wave excitations appear to be incompatible with a hard
Ising-like (1,1,1) anisotropy for the spins, and require continuous spin
degrees of freedom.  Similar issues had arisen in understanding the
``checkerboard" pattern of diffuse scattering in H=0 \cite{Gardner01, 
Kao}.  

\begin{figure}
\centering
\includegraphics[width=8.25cm]{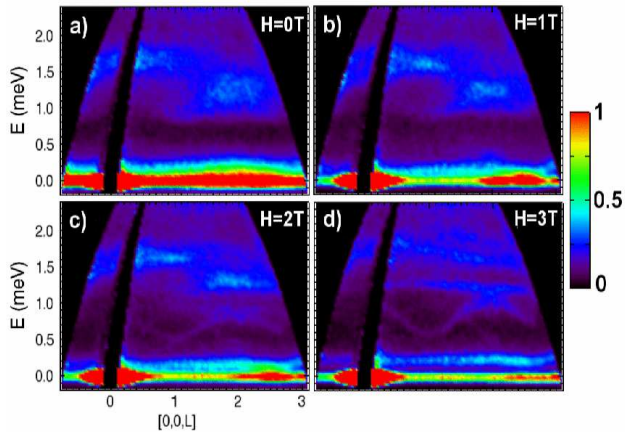}
\caption{Neutron scattering data within the (0,0,L) - energy plane at   
T=0.4 K and a) H=0, b) H=1 T, c) H=2 T, and d) H=3 T are shown.}
\label{Figure 4}
\end{figure}

The detailed magnetic structure of the high field ordered phase will be
reported on separately.  However, as seen in Fig. 1c) new Bragg peaks are
seen at all (H,H,L) indices within the field of view except (0,0,1).  
This pattern of observed reflections, although not the detailed relative
intensities, is similar to the low field antiferromagnetic state
reported by Mirebeau et al. \cite{Mirebeau05, MirebeauPF} in {\tto} 
under application of both a uniaxial and a
hydrostatic applied pressure.  This suggests that these two long range
ordered states are related, and that the appearance of an ordered state 
at high fields under ambient pressure may be due to strong
magnetoelastic effects, as have been reported in {\tto} \cite{Russians}.

We hope these results for a new phase diagram for {\tto} at ambient 
pressure, and the results for the spin wave spectrum within the 
magnetically ordered phase motivate a complete understanding of the 
complex and exotic ground state of this geometrically frustrated magnet.

We wish to acknowledge useful contributions from M.J.P. Gingras, G. 
Luke and J. Rodriguez. This work was supported by NSERC of
Canada, and utilized facilities supported in part by the NSF under
Agreements DMR-9986442 and DMR-0086210 and by the U.S. Department of
Energy under contracts DE-AC02-98CH10886.


%
%






\end{document}